# Nanofluid Filled Enclosures: Potential Photo-thermal Energy Conversion and Sensible Heat Storage Devices

Inderpreet Singh[a], Satbir Singh Sehgal[a], and Vikrant Khullar[b,]*

[a]Mechanical Engineering Department, Chandigarh University, Gharuan, Mohali-140413, Punjab, India
[b]Mechanical Engineering Department, Thapar Institute of Engineering & Technology, Patiala-147004, Punjab, India

*Corresponding author. Email address: vikrant.khullar@thapar.edu

**ABSTRACT**

In the present work we propose "nanofluid filled enclosures" as potential photo-thermal energy conversion and sensible heat storage devices. Herein, the optical charging of the nanofluid has been modeled as "solar radiant energy - nanoparticles" interaction. The subsequent energy redistribution has been modeled as coupled transport phenomena involving mass, momentum and energy transport. In particular, nanofluid filled enclosure with adiabatic, and isothermal (through convective) boundaries have been analyzed to decipher the fundamental limits of sensible heat storage and thermal discharging capacities respectively. Furthermore, the effect of nanoparticles volume fraction on the photo-thermal energy conversion mechanisms and its redistribution thereof has been critically investigated. Detailed analysis reveals that under similar operating conditions, in volumetric absorption mode (i.e., at low nanoparticles volume fraction) nanofluid filled enclosure has higher sensible heat storage (7% - 27% higher) and thermal discharging (16% - 47% higher) capacities than in the corresponding surface absorption mode (i.e., at high nanoparticles volume fraction). Overall, "nanofluid filled enclosures", particularly in volumetric absorption mode, could be deployed for efficient solar thermal conversion and storage.

## 1. INTRODUCTION

In the backdrop of the ever increasing demand for energy, coupled with reducing fossil fuel reserves and increasing climate disruption; researchers are extensively working towards improving the incumbent renewable energy based technologies. Undoubtedly, solar energy has emerged as potentially the most viable option among the available renewable energy resources; particularly, the solar photovoltaic technologies. However, the solar thermal energy conversion route (which presents a more practical solution for heating and cooling applications, and accounts for nearly 50% of the global energy consumption [1]) is yet to witness widespread deployment owing to lower conversion efficiencies and initial capital investments. In this realm, the present state - of - the - art technologies employ solar selective surfaces for the photo-thermal energy conversion process [2]. Owing to inherent thermal barrier between the absorbing surface and the working fluid; there is significant "temperature overheat"; particularly at high flux conditions - resulting in appreciable performance reduction [3-6].

In this regard, volumetric absorption through nanofluids presents an alternative route to effectively convert the incident solar energy to the thermal energy of the liquid which could be directly put to use or could be stored for non-sunshine hours. Herein, the nanoparticles dispersed in the liquid absorb the incident solar energy and subsequently transfers it to the surrounding liquid. Particle size being very small; negligibly small temperature difference exists between the nanoparticle and the surrounding liquid - hence resulting in efficient photo-thermal energy conversion [7-13]. In order to realize efficient solar thermal energy conversion platforms based on volumetric absorption principle, a detailed investigation into transport mechanisms involved in such systems is warranted. In this direction, to decipher the fundamental performance limits, researchers across the globe have modeled nanofluid based volumetric absorption receivers as nanofluid filled enclosures irradiated from top [3, 4, 8, 9, 14-19]. Figure 1 presents a list of selected works pertinent to the analysis of heat transfer mechanisms involved in nanofluid filled enclosures. Critical reviews into the aforementioned reported works reveals that most of the studies (whether theoretical or experimental) focus on determining temperature distribution within the enclosure without regarding natural convection effects. Moreover, the analysis has been limited to nanofluid enclosures with "adiabatic" boundaries. Furthermore, the intricate coupled interactions between the governing transport mechanisms under various operating conditions have not been dealt with in a comprehensive and holistic manner.

In the present work, we propose that "nanofluid filled enclosures" could be employed as "sensible heat storage devices" - involving optical charging and thermal discharging processes. Furthermore, the present work serves to identify the predominant operational heat transfer mechanisms (and quantify their magnitudes) in volumetrically absorbing photo-thermal energy conversion systems. Finally, an attempt has been made to investigate the effects of nanoparticles concentration, boundary conditions (adiabatic, convective and isothermal) and incident flux on the overall performance of such systems.

## 2. MODELING OPTICAL CHARGING AND THERMAL DISCHARGING PROCESSES IN NANOFLUID FILLED ENCLOSURES

### 2.1 Basic concept (of optical charging and thermal discharging processes) and constructional details

Optical charging essentially refers to interaction of incident solar energy with the nanofluid (through absorption and scattering mechanisms) and subsequent redistribution of the absorbed energy within the enclosure through conduction and advection mechanisms. Amorphous carbon nanoparticles dispersed in silicon oil has been considered as the representative nanofluid in the present work.

Thermal discharging refers to extracting sensible heat from the nanofluid filled enclosure by making a secondary heat transfer fluid flow around the left, bottom and right walls of the enclosure. Isothermal boundaries essentially represent the case when the convective heat transfer coefficient of the aforementioned secondary working fluid reaches a very high value.

Figure 2 shows constructional details of the two configurations of nanofluid filled enclosure studied in the present work. Configuration with adiabatic boundaries (see Fig. 2(a)) represents the case of optical charging without thermal discharging (except for the thermal losses from the top surface). Herein, we are interested in determining the limits of sensible heat storage capacity; i.e., assessing the fraction of incident energy that could be stored as sensible heat in the nanofluid. On the other hand, nanofluid filled enclosure with convective or isothermal boundaries (see Fig. 2(b)) simulate simultaneous optical charging with thermal discharging; i.e., estimating the fraction of incident energy that could be extracted from the enclosure as useful sensible heat under steady state conditions.

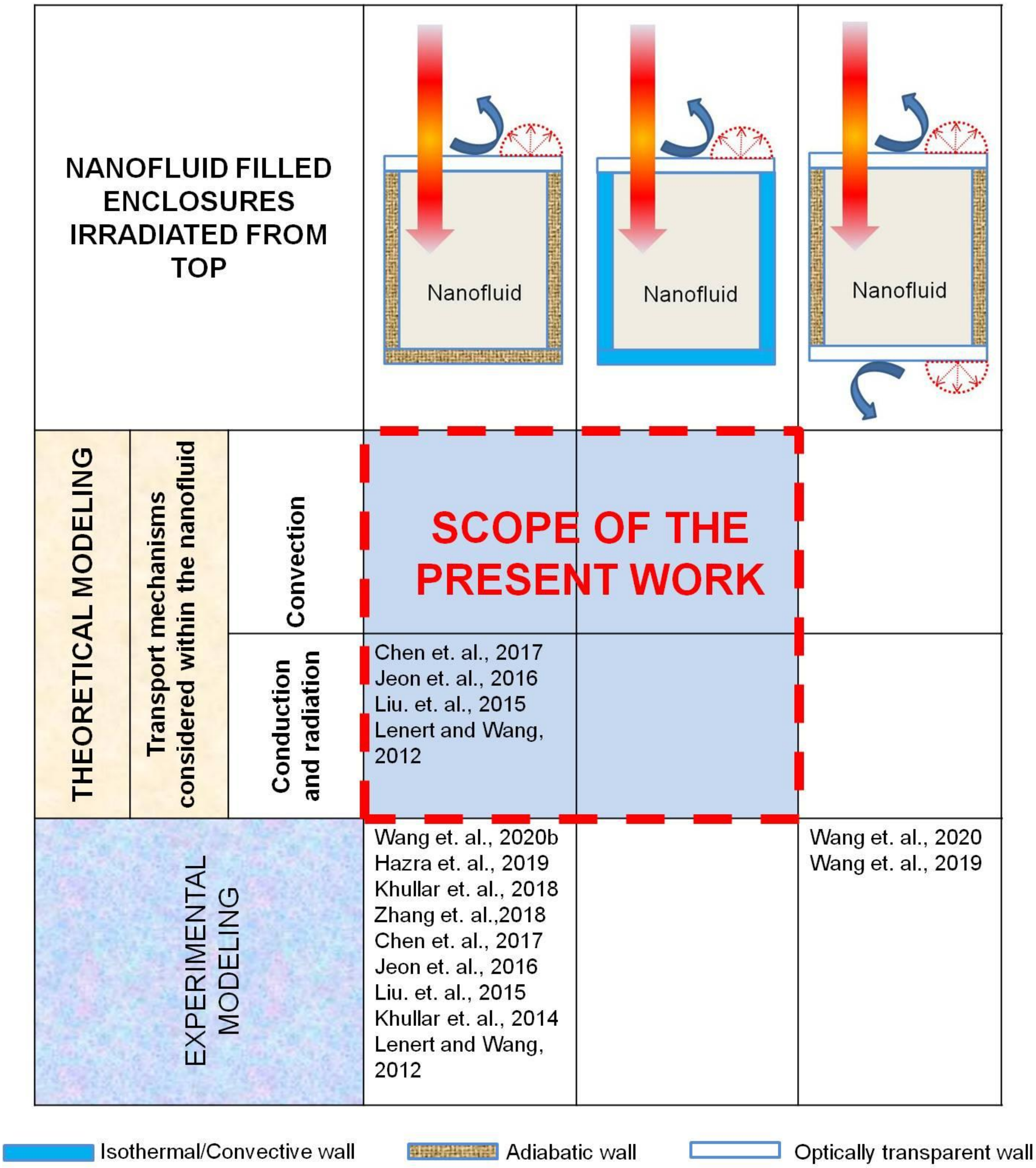

Fig. 1 Selected works pertinent to nanofluid filled enclosures heated from the top.

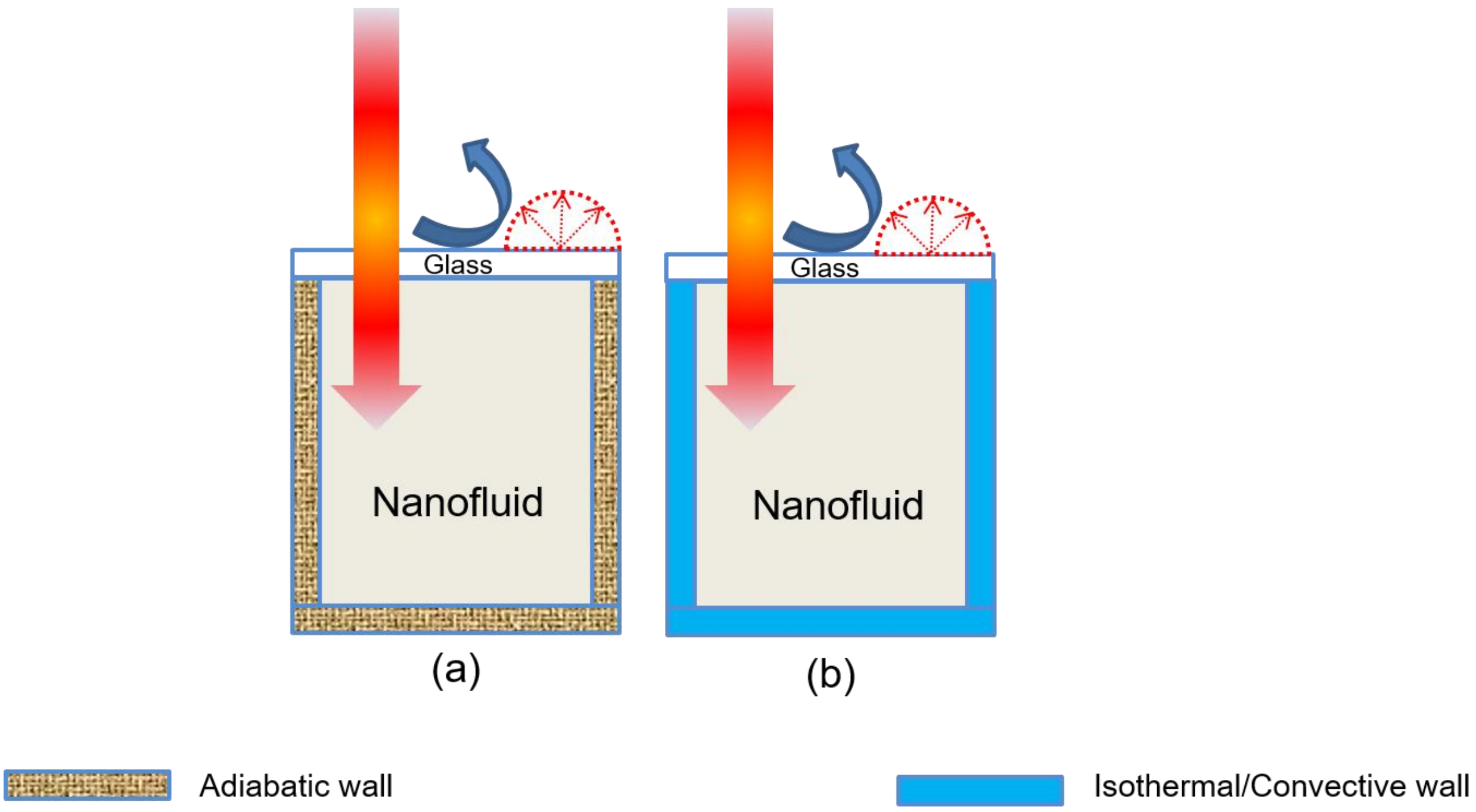

Fig. 2 Schematic representation showing the two nanofluid filled enclosure configurations studied in the present work; (a) adiabatic boundaries - simulating optical charging without thermal discharging, and (b) isothermal/convective boundaries - simulating optical charging with thermal discharging.

## 2.2 Theoretical modeling framework

***Underlying assumptions***

- Incident flux is assumed to strike normally at the top surface of enclosure [3].
- Bottom surface of the enclosure is considered to be perfectly reflecting; i.e., it reflects all the solar radiation that reaches the bottom surface of enclosure [3].
- Nanofluid has been modeled as single phase semitransparent participating media [3-7].
- Thermophysical properties of nanofluid have been assumed to be same as that of basefluid [3-7].

***Modeling sunlight-nanofluid interaction within the enclosure***

During passage of radiation (sunlight) through nanofluid along the depth direction of the enclosure, the magnitude of radiation changes due to absorption and scattering phenomena. Nanoparticles suspended in basefluid being very small compared to incident light wavelength lends scattering to be approximated as 'Rayleigh' and the predominant mode of radiation attenuation happens via non-radiative decay mechanism, i.e., absorption.
Absorption phenomenon in case of basefluid is mathematically represented as [20, 21]:

$$K_{abs,\lambda,bf} = \frac{4\pi\kappa}{\lambda}, \qquad (1)$$

where '$K_{abs,\lambda,bf}$' is the spectral absorption coefficient, '$\kappa$' is the spectral index of absorption, and '$\lambda$' is the wavelength.
In case of nanoparticles, attenuation of the incident radiation happens through absorption as well as scattering mechanisms. Both these mechanisms can be quantified by defining extinction coefficient and is mathematically given as [20 - 22]

$$K_{ext,\lambda,np} = \frac{1.5 f_v Q_{ext,\lambda}(\acute{\eta}, m)}{d}, \tag{2}$$

where, '$K_{ext,\lambda,np}$' is the spectral extinction coefficient, '$f_v$' is the volume fraction, '$d$' is hydrodynamic diameter (= 30nm in the present work), '$\acute{\eta}$' is the size parameter (= $\pi d/\lambda$), '$m$' is the normalized refractive index (= $m_p/n_f$), '$Q_{ext,\lambda}$' is the extinction efficiency of the particles such that [20 - 22]

$$Q_{ext,\lambda} = 4\acute{\eta}\,\mathrm{Im}\left\{\frac{m^2-1}{m^2+2}\left[1+\frac{\acute{\eta}^2}{15}\left(\frac{m^2-1}{m^2+2}\right)\frac{m^4+27m^2+38}{2m^2+3}\right]\right\}+\frac{8}{3}\acute{\eta}^4\,\mathrm{Re}\left\{\left(\frac{m^2-1}{m^2+2}\right)^2\right\} \tag{3}$$

Values of spectral index of absorption i.e '$\kappa$' for base-fluid (silicon oil) were obtained from Ref. [22]. For nanoparticles (amorphous carbon), values of spectral index of absorption, i.e., '$\kappa$' have been taken from Ref. [23].
After the complete calculation of extinction coefficient of nanoparticles and basefluid individually, the next step is to add both values to get final value of extinction coefficient for nanofluid as shown below

$$K_{ext,\lambda,nf} = K_{abs,\lambda,bf} + K_{\mathrm{ext},\lambda,np}, \tag{4}$$

***Modeling transport phenomena responsible for energy redistribution within the enclosure***
To model mass and momentum transport in the nanofluid filled enclosure; continuity (Eq. (5)) and Navier-Stokes (Eqs. (6), and (7)) equations need to be solved.

$$\frac{\partial(\rho u_{nf})}{\partial x} + \frac{\partial(\rho v_{nf})}{\partial y} = 0 \tag{5}$$

$$\frac{\partial(\rho u_{nf})}{\partial t} + \frac{\partial(\rho u_{nf} u_{nf})}{\partial x} + \frac{\partial(\rho v_{nf} u_{nf})}{\partial y} = -\frac{\partial p_{nf}}{\partial x} + \mu_{visc}\left(\frac{\partial^2 u_{nf}}{\partial x^2} + \frac{\partial^2 u_{nf}}{\partial y^2}\right) \tag{6}$$

$$\frac{\partial(\rho v_{nf})}{\partial t} + \frac{\partial(\rho u_{nf} v_{nf})}{\partial x} + \frac{\partial(\rho v_{nf} v_{nf})}{\partial y} = -\frac{\partial p_{nf}}{\partial y} + \mu_{visc}\left(\frac{\partial^2 v_{nf}}{\partial x^2} + \frac{\partial^2 v_{nf}}{\partial y^2}\right) + S_b \tag{7}$$

where $u_{nf}$ and $v_{nf}$ are the nanofluid velocities in $x$ and $y$ directions respectively; $\mu_{visc}$ is the dynamic viscosity of the nanofluid; $p$ is an effective pressure; $S_b$ is the Boussinesq source term and is mathematically given by Eq. (8) as

$$S_b = \rho_{ref}\, g\beta(T_{nf} - T_{ref}) \tag{8}$$

where '$\beta$', '$\rho_{ref}$', '$T_{nf}$' are the coefficient of thermal expansion, reference density and the local temperature of the nanofluid (found through solving energy equation; two way coupling exists between velocity and temperature fields) respectively; '$g$' is acceleration due to gravity, and '$T_{ref}$' is the reference temperature. The aforementioned governing differential equations are subjected to the following boundary conditions:

$$u_{nf} = v_{nf} = 0, \text{ for } x = 0 \text{ and } x = D \tag{9}$$

$$u_{nf} = v_{nf} = 0, \text{ for } y = 0 \text{ and } y = H \tag{10}$$

Furthermore, the temperature field within the nanofluid is found through solution of energy equation (Eq. (11))

$$\frac{\partial(\rho T_{nf})}{\partial t} + \frac{\partial(\rho u_{nf} T_{nf})}{\partial x} + \frac{\partial(\rho v_{nf} T_{nf})}{\partial y} = \frac{k_{nf}}{C_{ps,nf}}\left(\frac{\partial^2 T_{nf}}{\partial x^2} + \frac{\partial^2 T_{nf}}{\partial y^2}\right) + \frac{S^T}{C_{ps,nf}} \tag{11}$$

where $S^T$ is the radiative source term and is mathematically expressed as Eq. (12)

$$S^T = -\frac{\partial q_r}{\partial y} \tag{12}$$

where $q_r$ is the radiative flux (summed over all the incident wavelengths, 0.4$\mu$m - 2.5$\mu$m) and is mathematically defined as

$$q_r(y) = \sum_{\lambda=0.4\mu m}^{\lambda=2.5\mu m} I_\lambda \tag{13}$$

Herein, the intensities along the depth direction have been computed using one dimensional radiative transfer equation (RTE) and is described by Eq. (14) [3]

$$\frac{dI_\lambda}{dy} = -K_{\lambda,ext,nf} I_\lambda \tag{14}$$

where, '$I_\lambda$' is radiative intensity at particular wavelength denoted by '$\lambda$', '$K_{\lambda,ext,nf}$' is the spectral extinction coefficient of the nanofluid.
Solving RTE for the intensity values results in finding the values divergence of radiative flux and hence the source term ($S^T$) in energy equation (Eq. (11)). The temperature field within the enclosure is then determined through solution of the energy equation subject to the 'free convective and radiation' boundary condition at the top wall, mathematically this is given by Eq. (15) as

$$-k_{nf} a_{s_top} \frac{\partial T}{\partial y}\Big|_{y=0} = h_{t_conv} a_{s_top}(T_{s_top} - T_{amb}) + \sigma a_{s_top}(T^4_{s_top} - T^4_{amb}) = Q''_{loss_conv} + Q''_{loss_rad} \tag{15}$$

where, $h_{t_conv} = 0.27 Ra^{1/4} ........ 10^5 \le Ra \le 10^{10}, \mathrm{Pr} = 0.71$ [24]

Whereas, the following boundary conditions are at bottom and side walls:
Case 1: Adiabatic boundary condition

$$-k_{nf} a_{s_left} \frac{\partial T}{\partial x}\Big|_{x=0} = h_L a_{s_left}(T_{s_left} - T_{amb}) \tag{16a}$$

$$-k_{nf} a_{s_right} \frac{\partial T}{\partial x}\Big|_{x=0} = h_R a_{s_right} (T_{s_right} - T_{amb})$$

(16b)

$$-k_{nf} a_{s_bottom} \frac{\partial T}{\partial y}\Big|_{y=H} = h_B a_{s_bottom} (T_{s_bottom} - T_{atm}) \quad (16c)$$

where convective heat transfer coefficient for left, right and bottom walls, $h_L = h_R = h_B = 0$ Wm$^{-2}$K$^{-1}$

Case 2: Convective boundary condition
Herein, the Eqs. (16a) - (16c) are operational with convective heat transfer coefficient for left, right and bottom walls as:

a) $h_L = h_R = h_B = 100$ Wm$^{-2}$K$^{-1}$
b) $h_L = h_R = h_B = 1000$ Wm$^{-2}$K$^{-1}$
c) $h_L = h_R = h_B = 10000$ Wm$^{-2}$K$^{-1}$

The chosen typical heat transfer coefficient values correspond to the free and forced convection regimes [25, 26].

Case 3: Isothermal boundary condition

$$T_{nf}(y, H) = T_{amb} \quad (17a)$$

$$T_{nf}(x, 0) = T_{amb} \quad (17b)$$

$$T_{nf}(x, L) = T_{amb} \quad (17c)$$

## 2.3 Numerical modeling

The transport phenomena within the enclosure have been mathematically modeled through governing differential equations and the corresponding boundary conditions. Now, the next step is to solve these equations to get the temperature and flow fields within the nanofluid filled enclosure. Given the fact that the aforementioned equations are nonlinear coupled partial differential equations, numerical solution route has been sought. This essentially involves physical domain as well as equation discretization to compute temperatures, velocities and pressures at discrete point with the enclosure.

The physical domain (i.e. the nanofluid filled enclosure) has been discritized into finite number of finite sized control volumes. In particular, 'staggered grid' arrangement is employed to compute velocity *($u_{nf}$, $v_{nf}$)*, pressure *($p_{nf}$)* and temperature *($T_{nf}$)* for each control volume [27, 28].

The governing differential equations have been discritized using semi-implicit finite control volume approach. Essentially, SIMPLE algorithm [28 - 29] has been implemented in MATLAB. Figures 3(a) and 3(b) represent flow chart of the algorithm implemented and the typical values of thermophysical properties/parameters/boundary conditions employed in the present work.

***Grid independence test***

In order to come up with solutions which are independent of the grid size chosen, we compute mid-plane temperatures along receiver depth for various grid sizes (60 × 60, 80 × 80, 120 × 120, 140 × 140 and 160 × 160). Looking at Fig. 4 reveals that mid-plane temperature values for 120 × 120 grid size are very near to that for 140 × 140 and 160 × 160 grid sizes. Therefore, grid size of 120 × 120 control volumes is the preferable choice, considering computational time and cost.

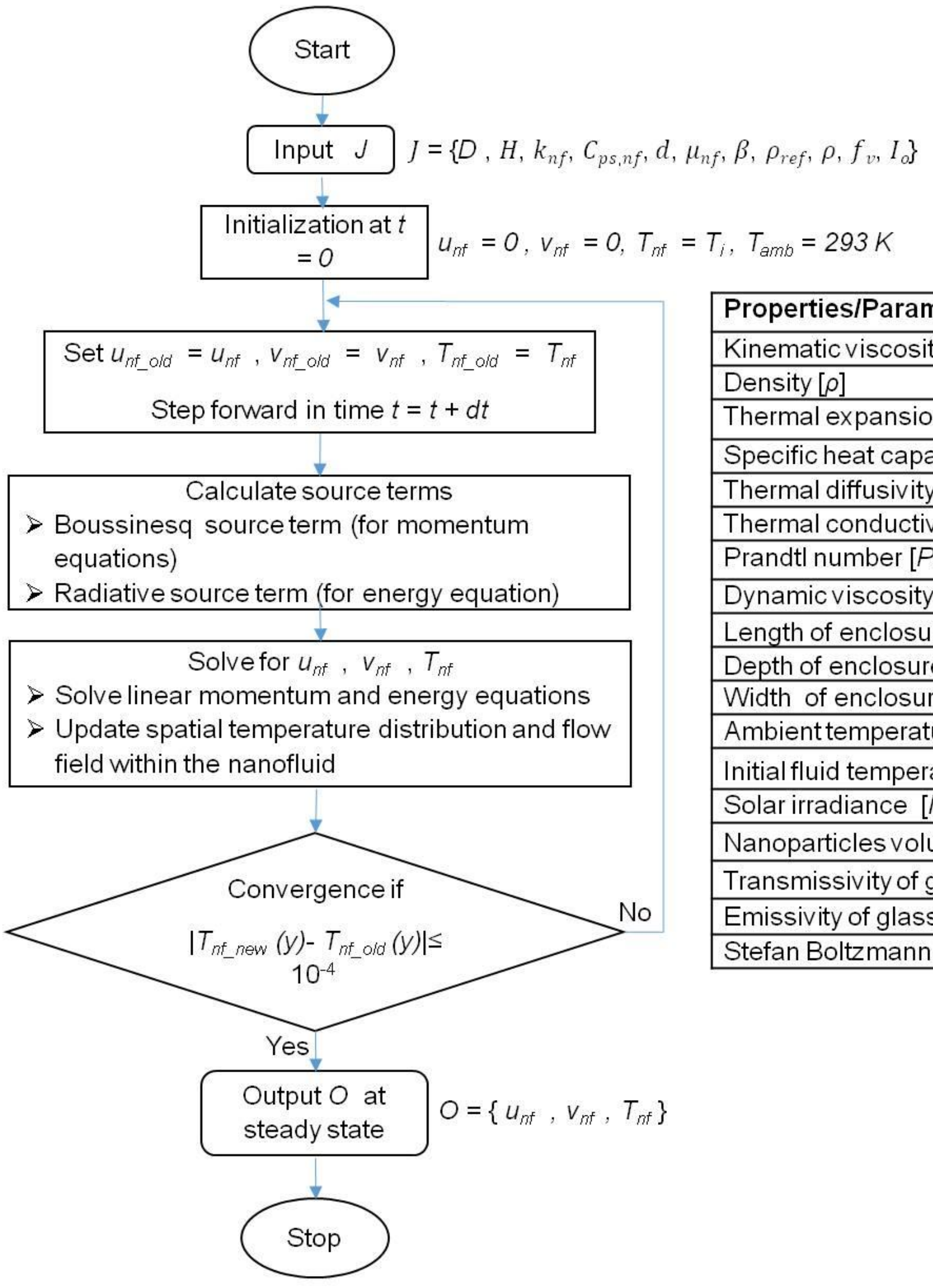

| Properties/Parameters | Values/Range |
|---|---|
| Kinematic viscosity [$\upsilon$] | $5\times10^{-6}$ $m^2s^{-1}$ |
| Density [$\rho$] | 910 $kgm^{-3}$ |
| Thermal expansion coefficient [$\beta$] | $1.1\times10^{-3}$ $K^{-1}$ |
| Specific heat capacity [$C_{ps}$] | 1400 $Jkg^{-1}K^{-1}$ |
| Thermal diffusivity [$\alpha$] | $8.6\times10^{-8}$ $m^2s^{-1}$ |
| Thermal conductivity [$k_{nf}$] | 0.11 $Wm^{-1}K^{-1}$ |
| Prandtl number [$Pr$] | 58 |
| Dynamic viscosity [$\mu$] | $10^{-3}$ $kgm^{-1}s^{-1}$ |
| Length of enclosure [$L$] | 1m |
| Depth of enclosure [$H$] | 0.04 m |
| Width of enclosure [$D$] | 0.04 m |
| Ambient temperature [$K$] | 293K |
| Initial fluid temperature [$T_i$] | 293K |
| Solar irradiance [$I_o$] | 1 Sun - 40 Suns |
| Nanoparticles volume fraction [$f_v$] | $10^{-1}$,$10^{-4}$, $5\times10^{-5}$ and $10^{-5}$ |
| Transmissivity of glass plate [$\tau$] | 0.95 |
| Emissivity of glass plate [$\varepsilon$] | 0.88 |
| Stefan Boltzmann constant [$\sigma$] | $5.67\times10^{-8}$ $Wm^{-2}K^{-4}$ |

(b)

Fig. 3 (a) Flow Chart showing the algorithm implemented to compute temperature and flow field, and (b) thermophysical properties [30]/parameters/boundary conditions employed in the present work.

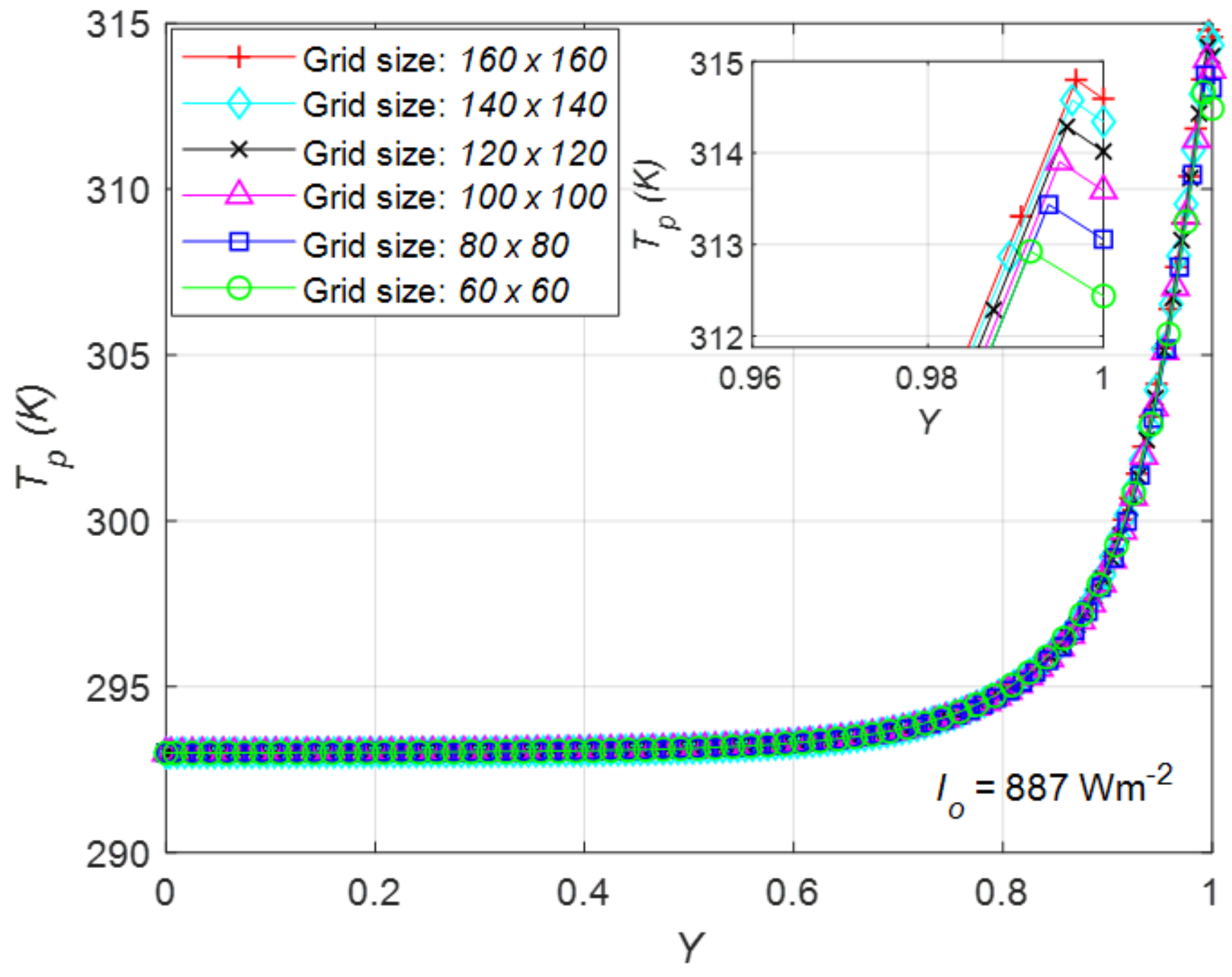

Fig. 4 Mid-plane temperatures along receiver depth for various grid sizes (60 × 60, 80 × 80, 120 × 120, 140 × 140 and 160 × 160).

***Model validation***

The developed theoretical modeling framework has been validated with experimental and numerical results reported by Wolff et al, 1988 [28] (for buoyancy driven flows, containing gallium and tin liquid metals in square cavity). Briefly, for validation a horizontal square cavity containing liquid gallium (with $Pr$ = 0.0208, $Ra = 1.682 \times 10^5$ ) as the working fluid was studied. Grid size of 50 × 50 has been used to validate the published results of Ref. [31] (see Fig. 5(a)). Furthermore, model has been validated with numerical results of Deshmukh et. al., 2011 [32] for higher Prandtl number fluid showing variation of dimensionless maximum temperature with Rayleigh number (see Fig. 5(b)). In both cases, results of present model are in fine agreement with results reported in Wolff et al., 1988 [31] and Deshmukh et. al., 2011 [32].

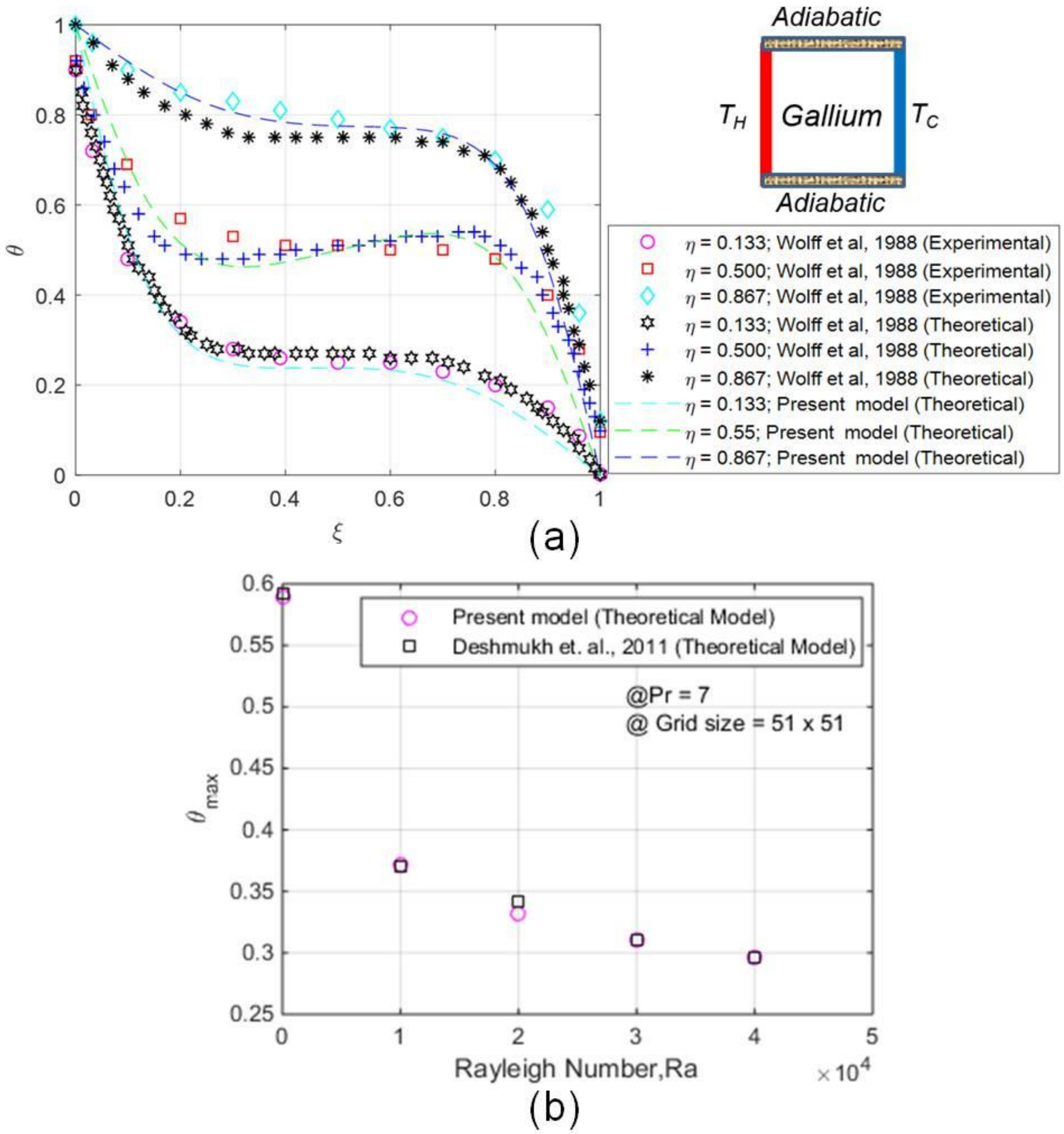

Fig. 5 (a) Comparison of the present theoretical modeling results with the experimental and theoretical results of Wolff et al, 1988 [31]; showing temperature profiles for liquid gallium, Pr = 0.0208, $Ra$ = 1.682 × $10^5$ and $\Delta T$ = 5$^o$C, and (b) comparison of current model with theoretical results of Deshmukh et. al., 2011[32] considering all four walls as isothermal (at 0K) with internal heat generation ($Q'$); showing maximum dimensionless temperature variation with Rayleigh number.

## 3. RESULTS AND DISCUSSION

### 3.1 Effect of nanoparticles volume fraction on intensity attenuation within the nanofluid filled enclosure

Figures 6(a) and 6(b) show the attenuation of incident solar radiation as it passes through the nanofluid in forward and backward (reflection from bottom surface) directions respectively. It is clearly apparent from the graphs that there is exponential decay of the incident radiation. Further, as the nanoparticles volume fraction increases, more and more radiation decay happens in first few top layers of the nanofluid. In other words, photo-thermal energy conversion mechanisms transforms from volumetric absorption mode to surface absorption mode as the nanoparticles volume fraction increases from $f_v = 10^{-5}$ to $10^{-1}$.

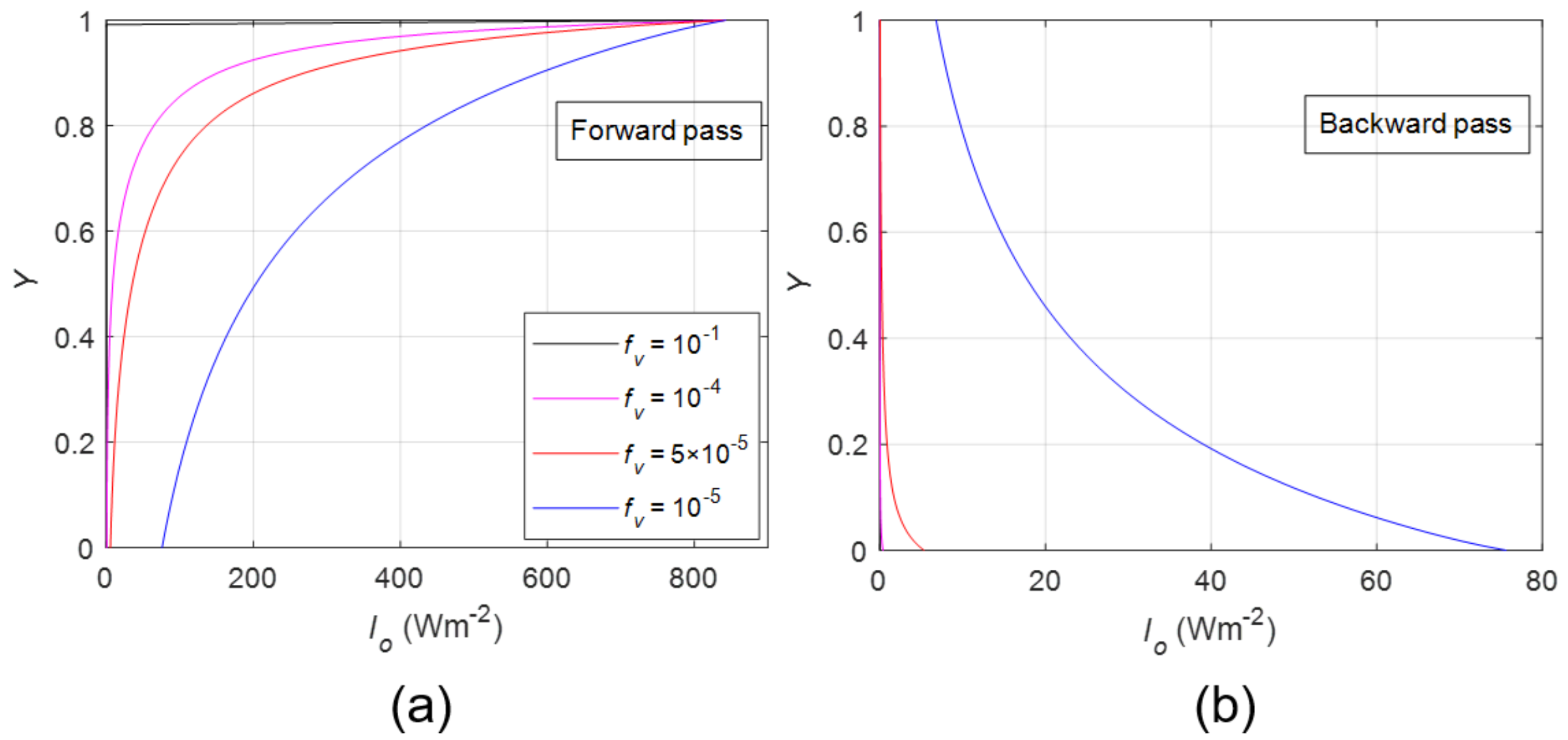

Fig. 6 Variation of incident solar intensity along depth direction for different nanoparticles volume fractions (a) forward pass, and (b) backward pass

**3.2 Assessing the fundamental limits of sensible heat storage capacity in nanofluid filled enclosures**

Adiabatic boundaries present the case of assessing the maximum possible solar thermal storage capacity. Figure 7 shows the temperature field within the nanofluid enclosure for various values of nanoparticles volume fraction. At high nanoparticles volume fraction ($f_v$ $=10^{-1}$), the energy is absorbed by the top nanofluid layers and the redistribution of this absorbed energy happens through conduction. On the other hand, at relatively lower nanoparticles volume fraction values, although still more energy absorbed in the top layers but also there is energy absorption in the middle and bottom layers - i.e., more distributed (volumetric) energy absorption. Furthermore, as there are thermal losses at the top; therefore the top nanofluid layer can become cooler that the adjacent layer below - resulting in the onset of natural convection. Moreover, the temperature field does not reach a steady state value - pointing that transition from laminar to turbulent situation.

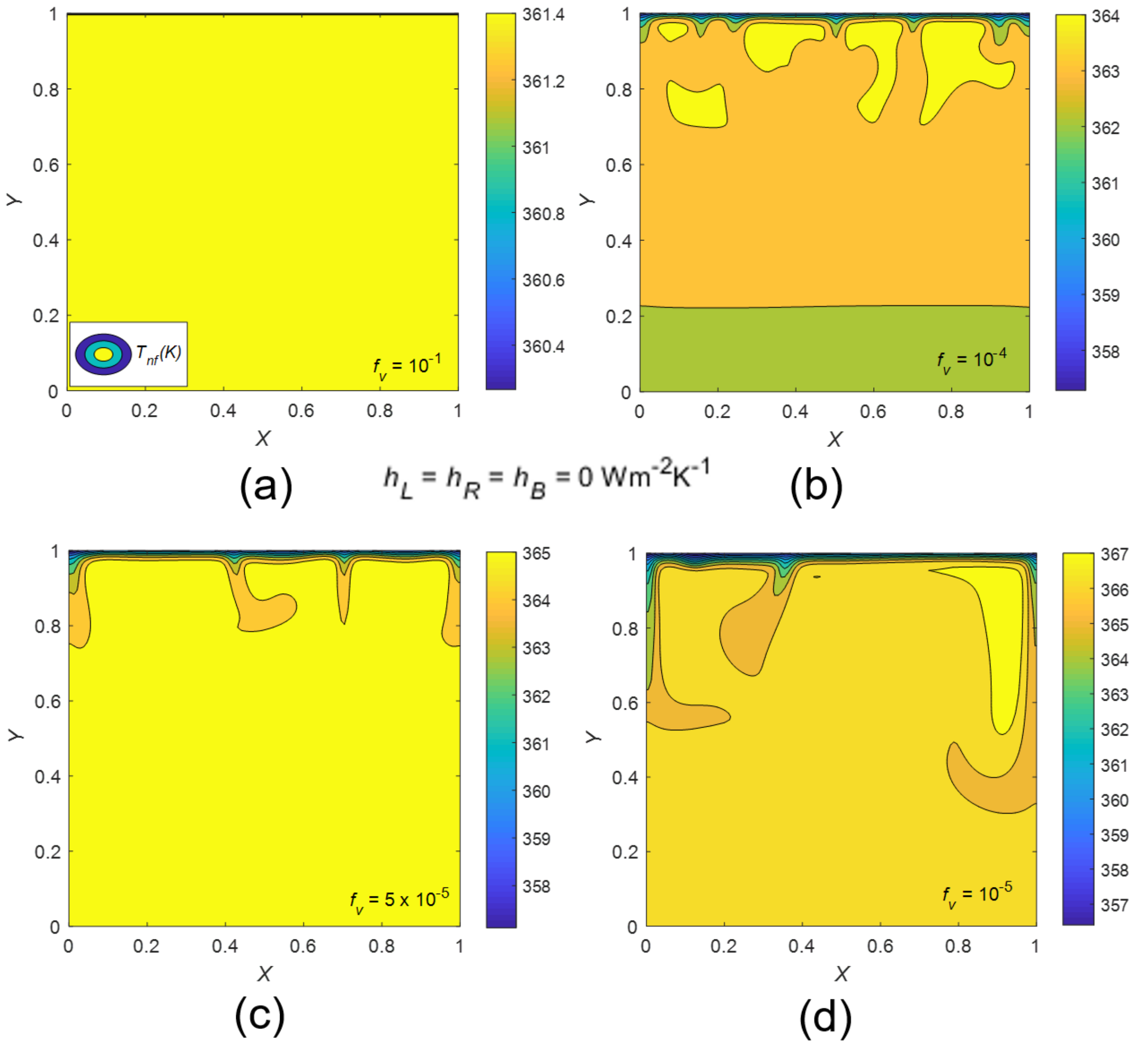

Fig. 7 Temperature distribution in a nanofluid filled enclosure operating at 1sun under adiabatic boundary condition ($h_L$ = $h_R$ = $h_B$ = 0 W m$^{-2}$K$^{-1}$) for (a) $f_v$ =10$^{-1}$ (b) $f_v$ =10$^{-4}$ (c) $f_v$ =5×10$^{-5}$, and (d) $f_v$ =10$^{-5}$.

Mid plane stream function values shown in Fig. 8(a) also confirms the fact that in surface absorption mode ($f_v$ =10$^{-1}$), the redistribution of absorbed energy essentially happens through conduction mechanism, whereas in volumetric absorption mode ($f_v$ = 10$^{-4}$, 5 ×10$^{-5}$ and 10$^{-5}$) both conduction and convection modes of heat transfer are operational. Moreover, convection intensifies as we reduce the nanoparticles volume fraction values - the stream function values being highest for $f_v$ = 10$^{-5}$. This effectively translates into higher average nanofluid temperatures in the enclosure (see Fig. 8(b)) - owing to reduced thermal losses in volumetric absorption mode.

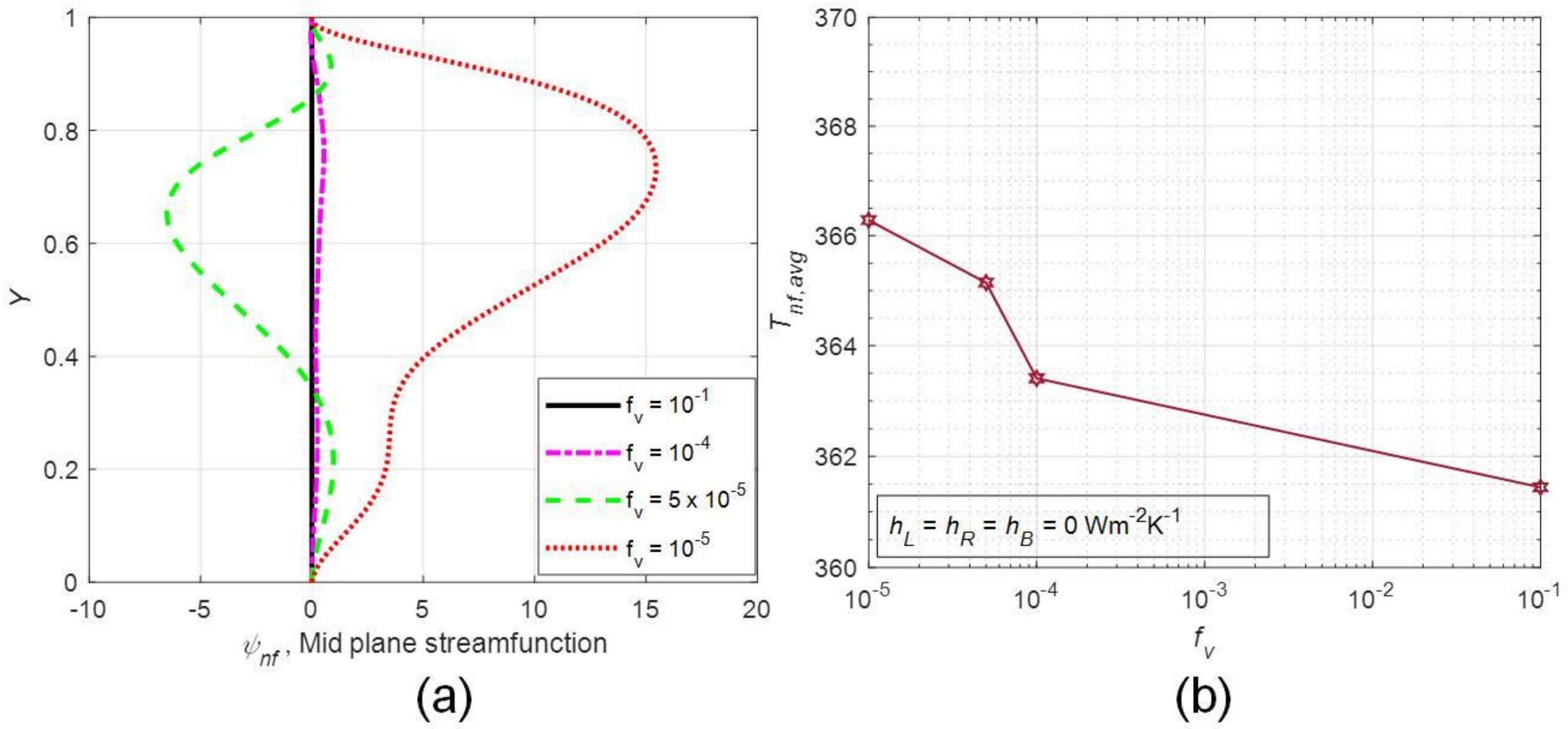

Fig. 8 (a) Mid-plane stream function values along the depth direction and (b) average nanofluid temperatures as a function of nanoparticles volume fractions.

In terms of sensible energy storage capacity; more energy could be stored in volumetric absorption mode and this benefit becomes more pronounced under high flux conditions (see Fig. 9).

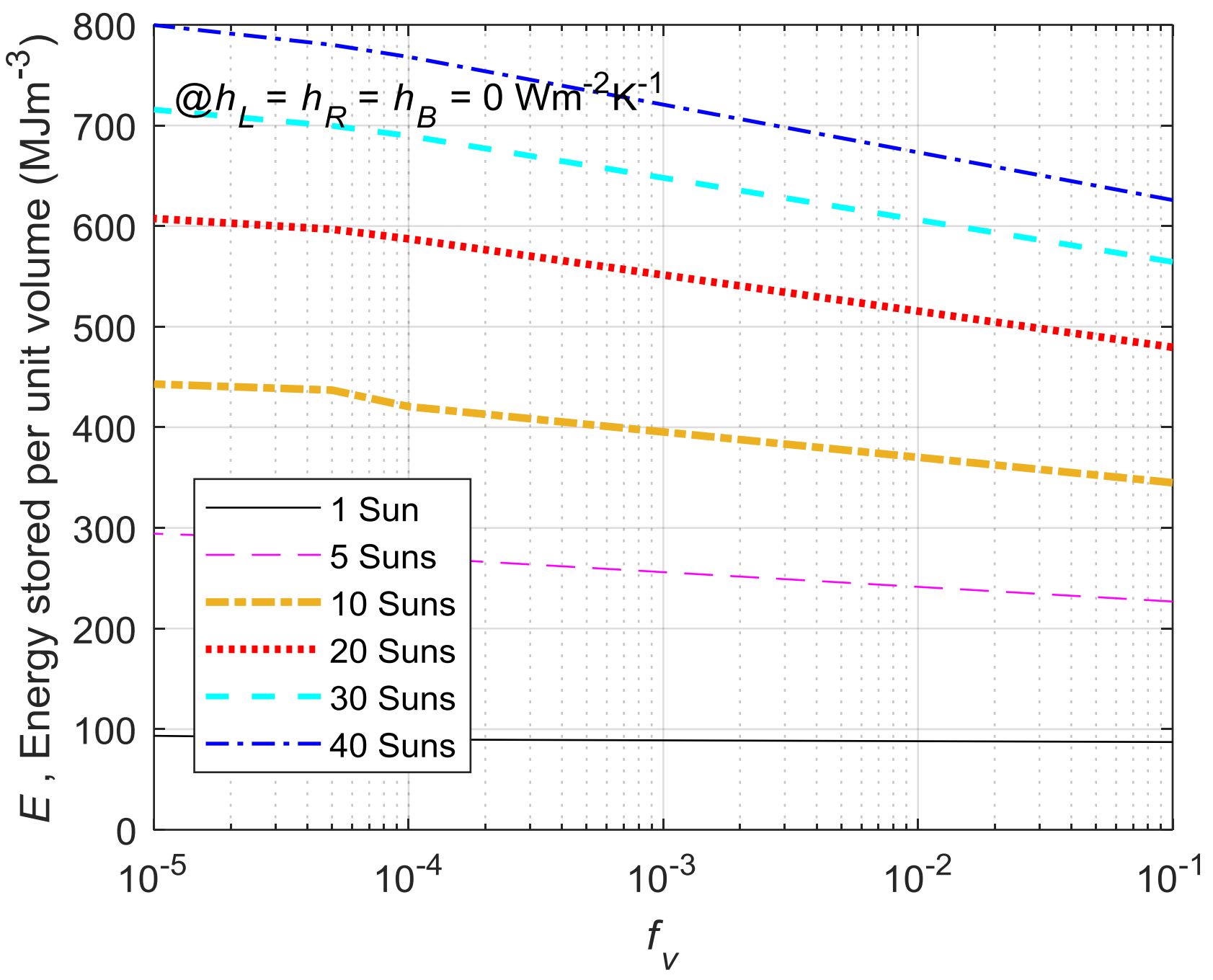

Fig. 9 Energy density storage capacity as a function of nanoparticles volume fraction.

### 3.3 Assessing the fundamental limits of thermal discharging in nanofluid filled enclosures

In order to extract sensible heat from the nanofluid enclosure, a secondary HTF needs to be circulated around the left, right and bottom boundaries. As the heat transfer coefficient value increases from 100Wm$^{-2}$K$^{-1}$ (through 1000Wm$^{-2}$K$^{-1}$ and 10000Wm$^{-2}$K$^{-1}$) to a very high value; the walls approximate isothermal boundary condition - resulting in more and more extraction of the thermal energy from the enclosure HTF. Figure 10 shows the temperature field for $h_L = h_R = h_B$ = 100Wm$^{-2}$K$^{-1}$, here, looking at the contours clearly points out that the absorbed energy redistribution within the nanofluid is through conduction mechanism irrespective of the mode of operation i.e., volumetric or surface absorption mode (independent of the nanoparticle volume fraction). Moreover, similar trend is followed for higher values of convective heat transfer coefficient (see Figs. 11 and 12) except for the fact that the solution convergence to steady state values becomes faster.

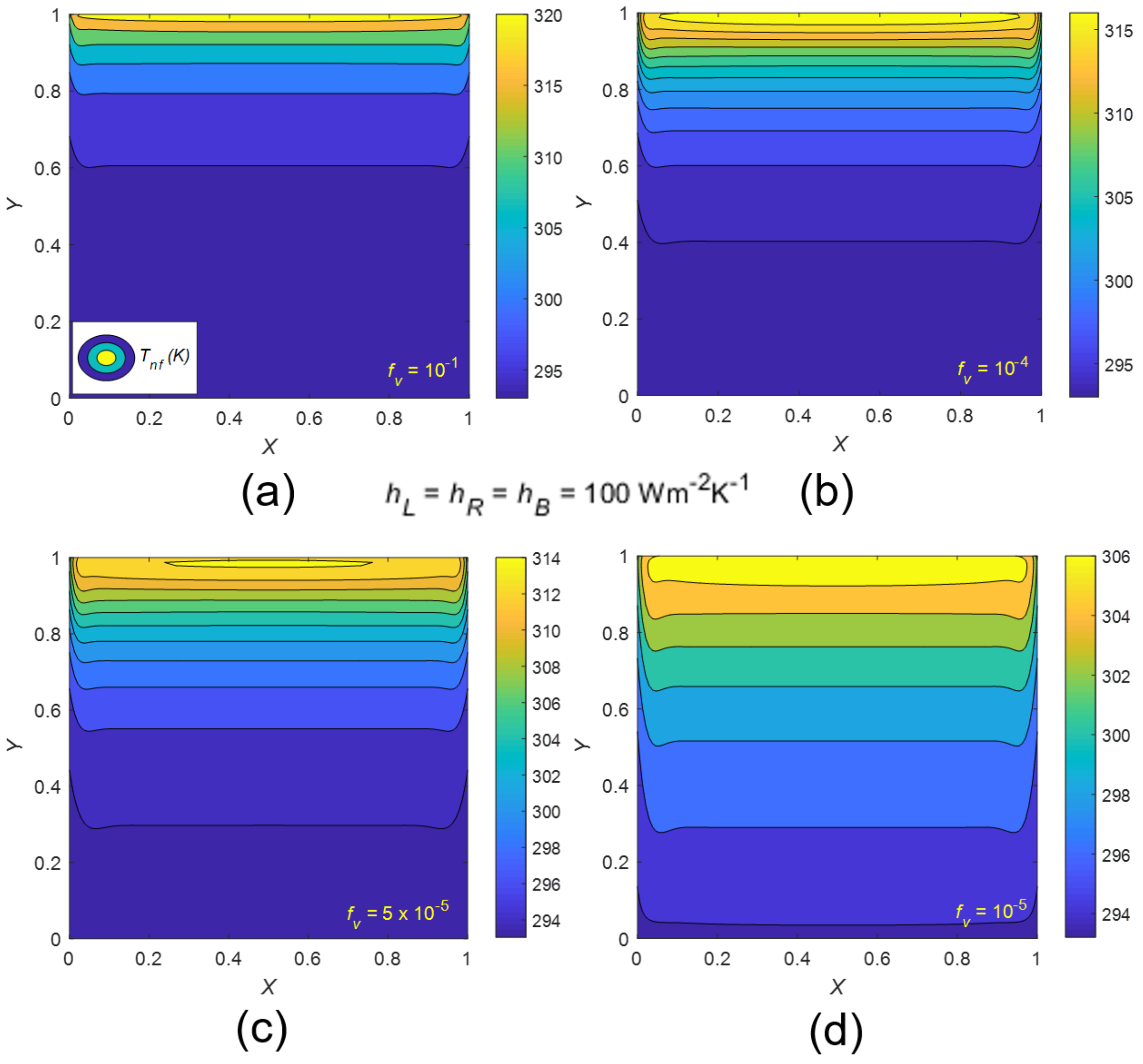

Fig. 10 Temperature distribution in nanofluid filled enclosure operating at 1sun under thermal discharging mode ($h$ = 100 W m$^{-2}$K$^{-1}$) for (a) $f_v$ = 10$^{-1}$, (b) $f_v$ = 10$^{-4}$, (c) $f_v$ = 5 × 10$^{-5}$ and (d) $f_v$ =10$^{-5}$.

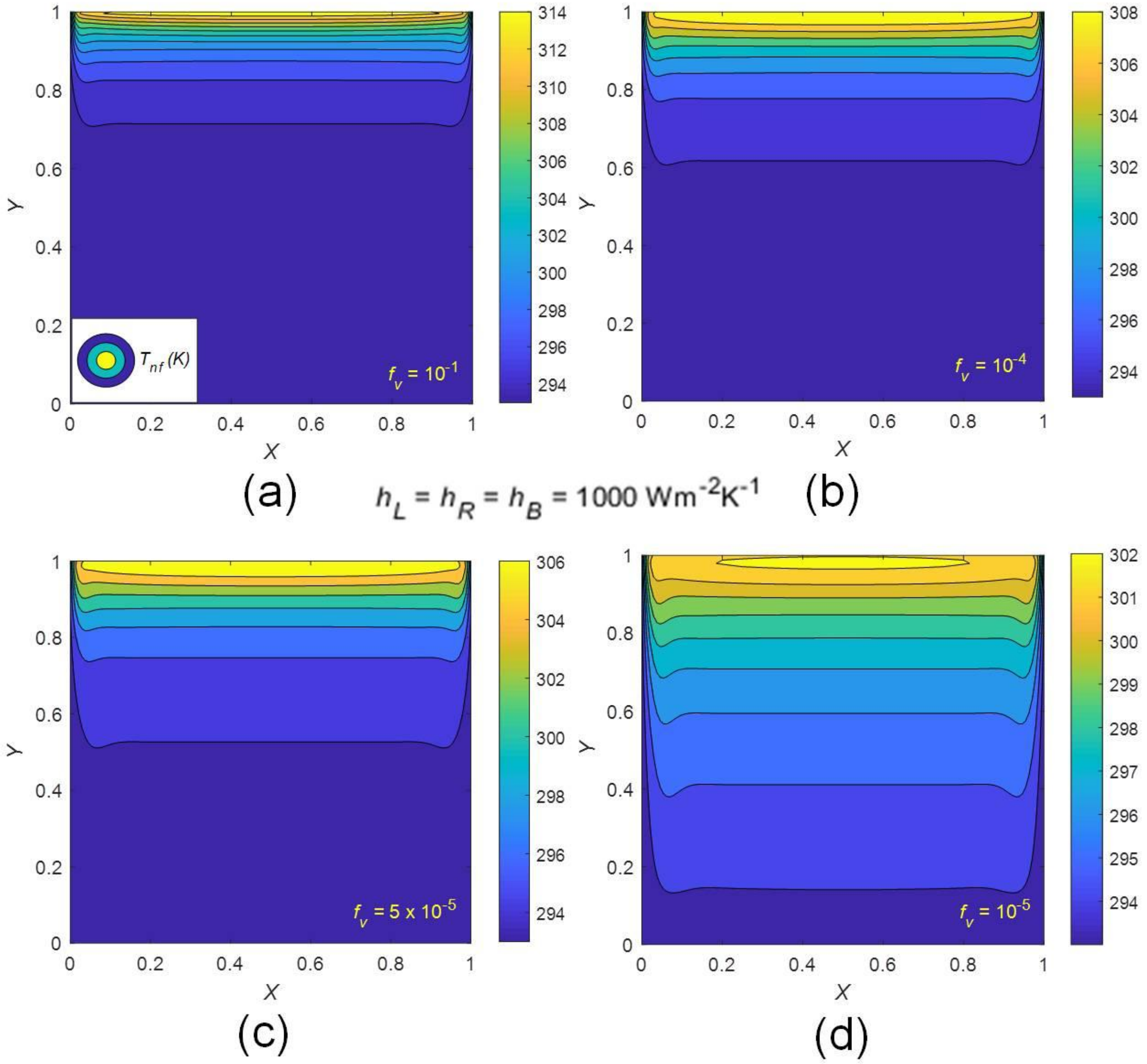

Fig. 11 Temperature distribution in nanofluid filled enclosure operating at 1sun under thermal discharging mode ($h$ = 1000 W m$^{-2}$K$^{-1}$) for (a) $f_v$ = 10$^{-1}$, (b) $f_v$ = 10$^{-4}$, (c) $f_v$ = 5 × 10$^{-5}$ and (d) $f_v$ =10$^{-5}$.

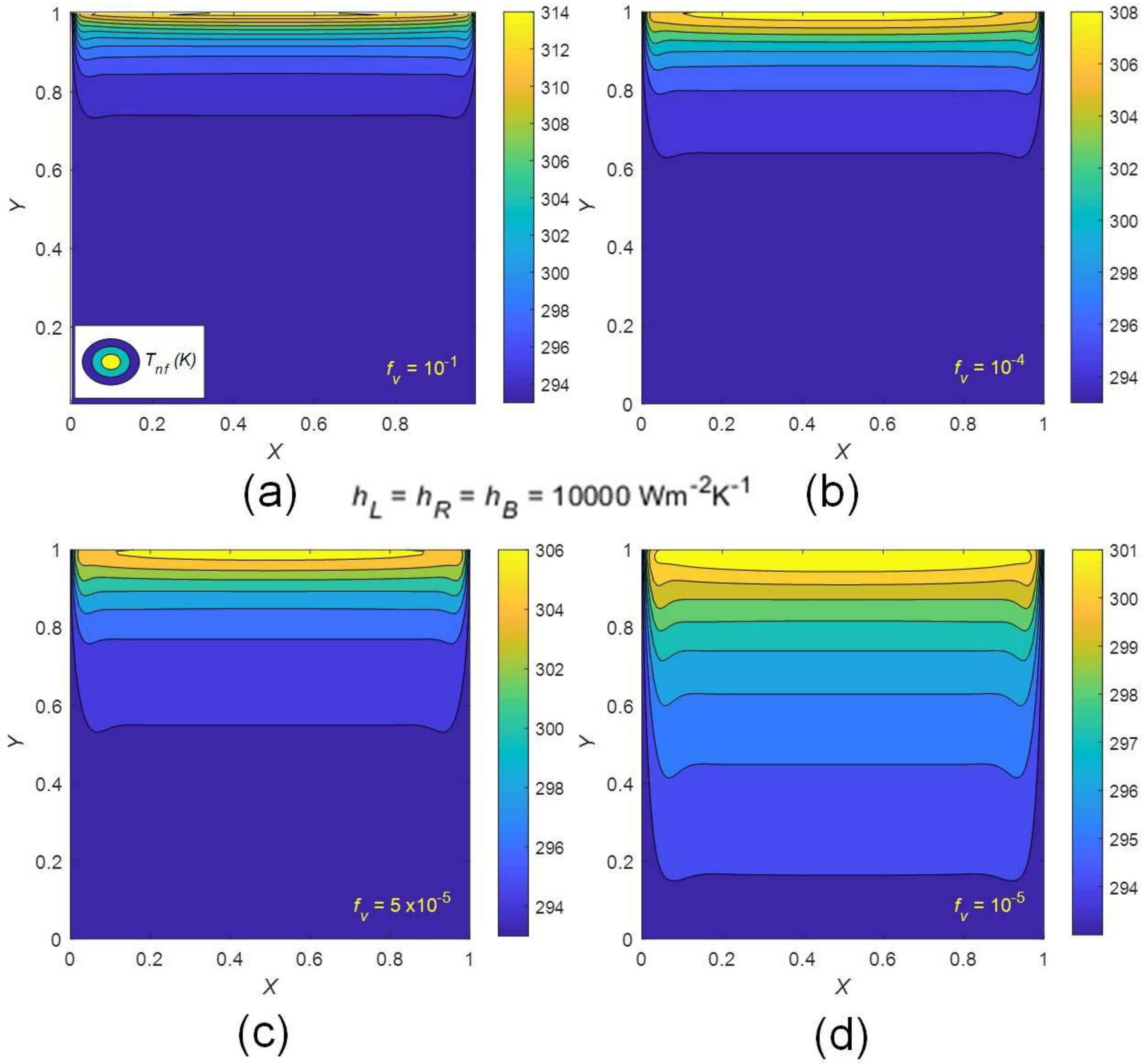

Fig. 12 Temperature distribution in nanofluid filled enclosure operating at 1sun under thermal discharging mode ($h$ = 10000 W m$^{-2}$K$^{-1}$) for (a) $f_v$ = 10$^{-1}$, (b) $f_v$ = 10$^{-4}$, (c) $f_v$ = 5 × 10$^{-5}$ and (d) $f_v$ =10$^{-5}$.

However, careful look into the mid plane temperatures (see Fig. 13) reveal that less pronounced temperature gradients are achieved in volumetric absorption mode - resulting in relatively lower losses from the top surface.

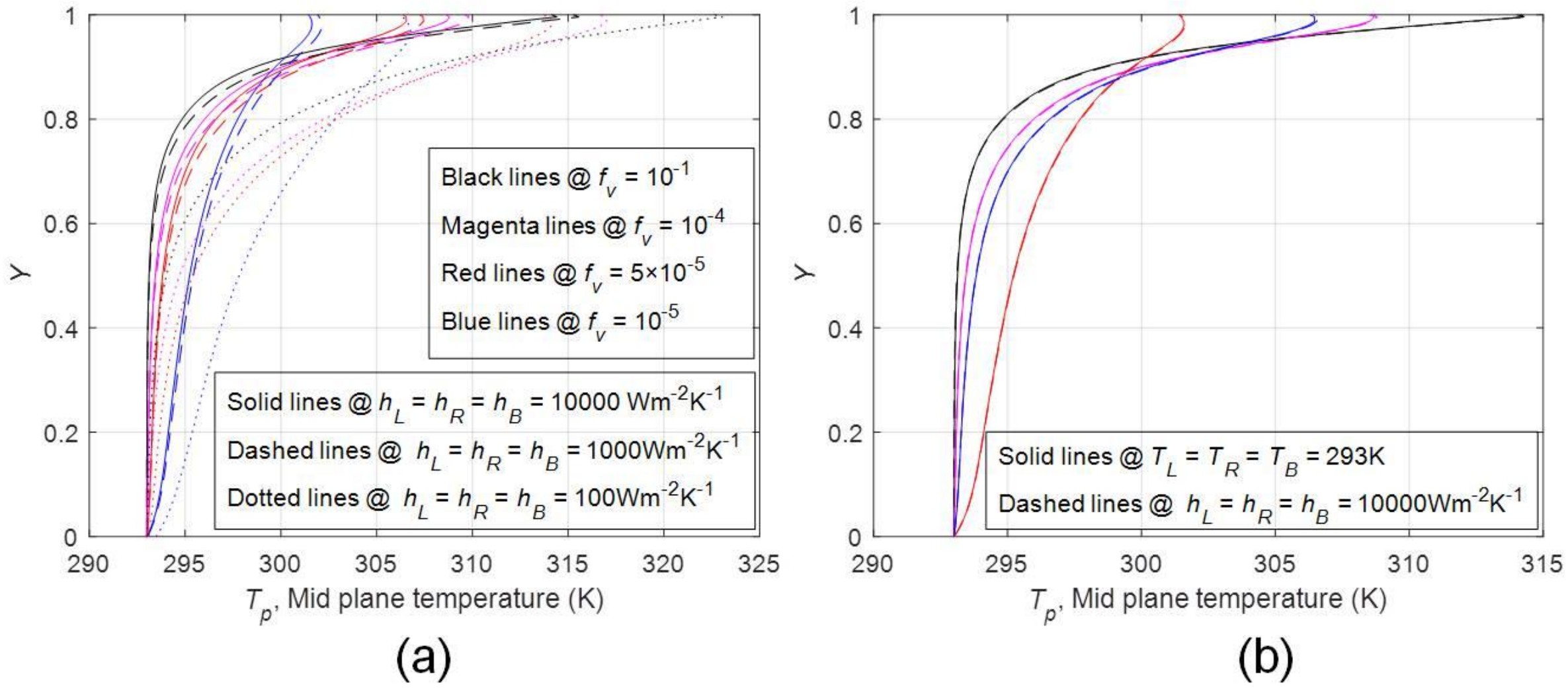

Fig. 13 Mid plane temperatures along enclosure depth direction as a function of nanoparticles volume fractions for (a) convective (100 $Wm^{-2}K^{-1}$,1000 $Wm^{-2}K^{-1}$,10000 $Wm^{-2}K^{-1}$) and (b) isothermal and convective (10000 $Wm^{-2}K^{-1}$) boundary conditions.

In terms of thermal discharging capacities also (see Fig. 14), volumetric absorption mode offers much higher extraction of the stored thermal energy from the enclosure. Furthermore, opposed to volumetric absorption, in surface absorption mode (i.e., at high nanoparticles volume fraction values), the thermal discharging capacity is a strong function of incident flux. This dependence becomes even more pronounced at lower values of convective heat transfer coefficient.

Mathematically, thermal discharging capacity is represented as:

$$1-\left[\frac{(Q''_{loss_conv}+Q''_{loss_rad})_{top_wall}}{\sum\limits_{all_walls} -k_{nf}\, a_s \frac{\partial T_{nf}}{\partial j}}\right]\times 100 \quad (18)$$

where $j=\begin{bmatrix} x \\ y \end{bmatrix}$, $x$ for left and right walls; while $y$ for bottom and top walls.

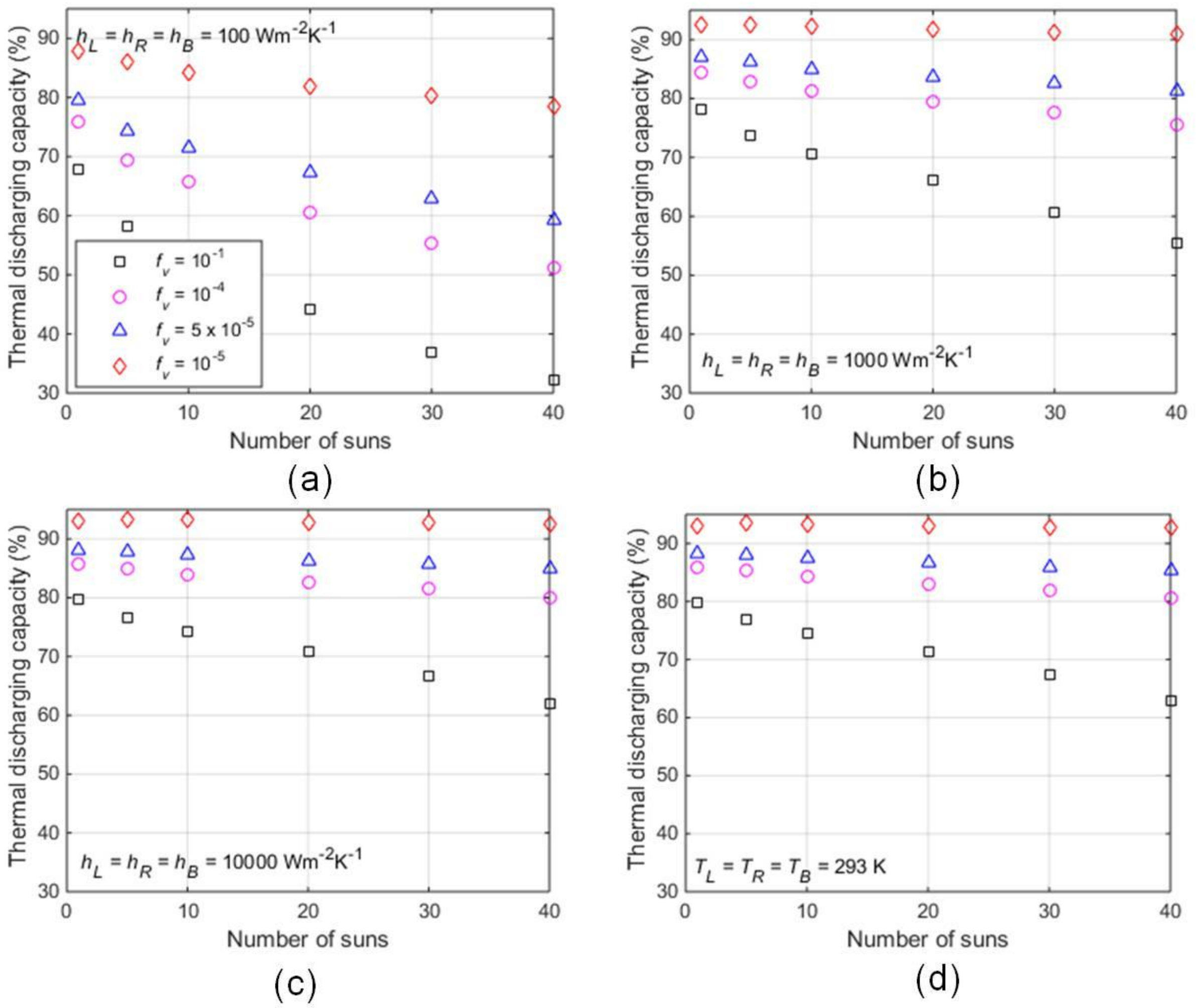

Fig. 14 Comparison of percentage thermal discharging capacities for various nanoparticles volume fractions under (a) convective (100 Wm$^{-2}$K$^{-1}$), (b) convective (1000 Wm$^{-2}$K$^{-1}$), (c) convective (10000 Wm$^{-2}$K$^{-1}$) and (d) isothermal boundary conditions.

## CONCLUSIONS

Through present work, fundamental performance limits of nanofluid filled enclosure as efficient photo-thermal and sensible storage devices have been determined. Furthermore, the present work serves to identify the predominant operational heat transfer mechanisms (and quantify their magnitudes) in surface/volumetrically absorbing photo-thermal energy conversion systems. Critical analysis reveals that careful control over nanoparticles volume fraction can significantly alter the transport mechanisms within the nanofluid and hence the performance characteristics. In surface absorption mode ($f_v$ =$10^{-1}$), the redistribution of absorbed energy essentially happens through conduction mechanism; whereas in volumetric absorption mode ($f_v$ = $10^{-4}$, 5 ×$10^{-5}$ and $10^{-5}$) both conduction and convection modes of heat transfer are operational. Moreover, this lends the nanofluid filled enclosure to have higher conversion efficiencies and storage capacities in volumetric absorption mode. Overall, nanofluid filled enclosures shall prove to be promising photo-thermal energy conversions platforms which could find application in a wide range of processes ranging from water heating, desalination, to electricity generation.

## ACKNOWLEDGEMENTS

This work is supported by DST-SERB (under Sanction order no. CRG/2021/003272). V.K. also acknowledges the support provided by Mechanical Engineering Department, Thapar Institute of Engineering & Technology Patiala, India. IS and SS acknowledge the support provided by Mechanical Engineering Department, Chandigarh University, Gharuan.

## NOMENCLATURE

***English Symbols:***

| | |
|---|---|
| *a* | surface area [$m^2$] |
| $C_{ps}$ | specific heat [J $kg^{-1}$ $K^{-1}$] |
| *d* | diameter of nanoparticles [nm] |
| *D* | width of enclosure [m] |
| *fv* | nanoparticles volume fraction |
| *g* | acceleration due to gravity [$ms^{-2}$] |
| *H* | height of enclosure [m] |
| *h* | convection heat transfer coefficient [$Wm^{-2}K^{-1}$] |
| *I* | solar irradiance [$Wm^{-2}$ $sr^{-1}$] |
| *K* | extinction coefficient |
| *k* | thermal conductivity [$Wm^{-1}K^{-1}$] |
| *L* | Length of enclosure [m] |
| *Pr* | Prandtl number |
| *Q* | Efficiency |
| *Q"* | total convective and radiative loss |
| *q* | radiative flux |
| *Ra* | Rayleigh number |
| *T* | local fluid temperature [K] |
| $T_p$ | mid-plane temperature [ K] |
| *u* | velocity component of fluid along x-direction [$ms^{-1}$] |
| *v* | velocity component of fluid along y- direction [$ms^{-1}$] |
| *x* | horizontal coordinate direction [m] |
| *y* | vertical coordinate direction [m] |
| *X* | dimensionless width of receiver, *x/L* |
| $\lambda$ | wavelength [µm] |
| υ | kinematic viscosity of nanofluid [$m^2$ $s^{-1}$] |
| $\rho$ | density of nanofluid [kg $m^{-3}$] |
| $\sigma$ | Stefan Boltzmann constant [$Wm^{-2}k^{-4}$] |
| $\tau$ | glass transmissivity |

***Subscripts:***

| | |
|---|---|
| *abs* | absorption |
| *amb* | ambient |
| *dy* | thickness of each control volume along *y* |
| *E* | control volume east of *P* |
| *e* | east face of control volume |
| *Ext* | Extinction |
| *N* | control volume north of *P* |
| *n* | north face of control volume |
| *nf* | Nanofluid |
| *Ref* | Reference |
| *S* | control volume south of *P* |
| *s* | south face of control volume |
| *s_bottom* | bottom surface |
| *s_left* | left wall surface |
| *s_right* | right wall surface |
| *t_rad* | top radiative loss |
| *s_right* | right wall surface |

| | | | |
|---|---|---|---|
| *Y* | dimensionless depth of receiver, *y/H* | *t_conv* | top convective loss |
| | | *visc* | Viscosity |
| | | *W* | control volume west of *P* |
| ***Greek symbols:*** | | *w* | west control volume |
| $\alpha$ | thermal diffusivity [$m^2s^{-1}$] | $\lambda$ | Spectral |
| $\beta$ | coefficient of volumetric expansion [$K^{-1}$] | | |
| $\varepsilon$ | emissivity of glass plate | | |
| $\gamma$ | size parameter | ***Superscript:*** | |
| $\theta$ | dimensionless temperature | *T* | term depicting temperature variable |
| $\kappa$ | spectral index of absorption | ***Abbreviations*** | |
| $\mu$ | dynamic viscosity [$kgm^{-1}s^{-1}$] | CV | control volume |
| $\Delta$ | temperature difference | | |

## REFERENCES

[1] IRENA, IEA, and REN21, Renewable Energy Policies in a Time of Transition: Heating and Cooling, no. November. 2020.

[2] Salvi, S. S., Bhalla, V., Taylor, R. A., Khullar, V., Otanicar, T. P., Phelan, P. E., & Tyagi, H. (2018). Technological advances to maximize solar collector energy output: a review. *Journal of Electronic Packaging, 140*(4), 040802.

[3] Lenert, A., & Wang, E. N. (2012). Optimization of nanofluid volumetric receivers for solar thermal energy conversion. *Solar Energy*, *86*(1), 253-265.

[4] Khullar, V., Tyagi, H., Hordy, N., Otanicar, T. P., Hewakuruppu, Y., Modi, P., & Taylor, R. A. (2014). Harvesting solar thermal energy through nanofluid-based volumetric absorption systems. *International Journal of Heat and Mass Transfer, 77*, 377-384.

[5] Khullar, V., Tyagi, H., Otanicar, T. P., Hewakuruppu, Y. L., & Taylor, R. A. (2018). Solar selective volumetric receivers for harnessing solar thermal energy. *Journal of Heat Transfer, 140*(6).

[6] Singh, A., Kumar, M., & Khullar, V. (2020). Comprehensive modeling, simulation and analysis of nanoparticles laden volumetric absorption based concentrating solar thermal systems in laminar flow regime. *Solar Energy, 211*, 31-54.

[7] Otanicar, T. P., Phelan, P. E., Prasher, R. S., Rosengarten, G., & Taylor, R. A. (2010). Nanofluid-based direct absorption solar collector. *Journal of renewable and sustainable energy*, *2*(3), 033102.

[8] Liu, J., Ye, Z., Zhang, L., Fang, X., & Zhang, Z. (2015). A combined numerical and experimental study on graphene/ionic liquid nanofluid based direct absorption solar collector. *Solar Energy Materials and Solar Cells*, *136*, 177-186.

[9] Jeon, J., Park, S., & Lee, B. J. (2016). Analysis on the performance of a flat-plate volumetric solar collector using blended plasmonic nanofluid. *Solar Energy*, *132*, 247-256.

[10] Freedman, J. P., Wang, H., & Prasher, R. S. (2018). Analysis of nanofluid-based parabolic trough collectors for solar thermal applications. *Journal of Solar Energy Engineering*, *140*(5).

[11] Singh, N., & Khullar, V. (2019). Efficient volumetric absorption solar thermal platforms employing thermally stable-solar selective nanofluids engineered from used engine oil. *Scientific reports*, *9*(1), 1-12.
[12] Singh, N., & Khullar, V. (2020). On-sun testing of volumetric absorption based concentrating solar collector employing carbon soot nanoparticles laden fluid. *Sustainable Energy Technologies and Assessments*, *42*, 100868.
[13] Singh, J., Mittal, M. K., & Khullar, V. (2021). Experimental Study of Single-Slope Solar Still Coupled With Nanofluid-Based Volumetric Absorption Solar Collector. *Journal of Solar Energy Engineering*, *144*(1), 011011.
[14] Chen, M., He, Y., Huang, J., & Zhu, J. (2017). Investigation into Au nanofluids for solar photothermal conversion. *International Journal of Heat and Mass Transfer*, *108*, 1894-1900.
[15] Zhang, R., Qu, J., Tian, M., Han, X., & Wang, Q. (2018). Efficiency improvement of a solar direct volumetric receiver utilizing aqueous suspensions of CuO. *International Journal of Energy Research*, *42*(7), 2456-2464.
[16] Hazra, S. K., Ghosh, S., & Nandi, T. K. (2019). Photo-thermal conversion characteristics of carbon black-ethylene glycol nanofluids for applications in direct absorption solar collectors. *Applied Thermal Engineering*, *163*, 114402.
[17] Wang, K., He, Y., Kan, A., Yu, W., Wang, D., Zhang, L., ... & She, X. (2019). Significant photothermal conversion enhancement of nanofluids induced by Rayleigh-Bénard convection for direct absorption solar collectors. *Applied Energy*, *254*, 113706.
[18] Wang, K., He, Y., Kan, A., Yu, W., Wang, L., Wang, D., ... & She, X. (2020). Enhancement of therminol-based nanofluids with reverse-irradiation for medium-temperature direct absorption solar collection. *Materials Today Energy*, *17*, 100480.
[19] Wang, K., He, Y., Liu, P., Kan, A., Zheng, Z., Wang, L., ... & Yu, W. (2020). Highly-efficient nanofluid-based direct absorption solar collector enhanced by reverse-irradiation for medium temperature applications. *Renewable Energy*, *159*, 652-662.
[20] Bohren, C. F., & Huffman, D. R. (2008). *Absorption and scattering of light by small particles*. John Wiley & Sons.
[21] Brewster, M. Q. (1992). *Thermal radiative transfer and properties*. John Wiley & Sons.
[22] Khullar, V., Bhalla, V., & Tyagi, H. (2018). Potential heat transfer fluids (nanofluids) for direct volumetric absorption-based solar thermal systems. *Journal of Thermal Science and Engineering Applications*, *10*(1)
[23] Stagg, B. J., & Charalampopoulos, T. T. (1993). Refractive indices of pyrolytic graphite, amorphous carbon, and flame soot in the temperature range 25 to 600 C. *Combustion and flame*, *94*(4), 381-396.
[24] Incropera, F. P., DeWitt, D. P., Bergman, T. L., & Lavine, A. S. (1996). *Fundamentals of heat and mass transfer* (Vol. 6). Wiley.
[25] Cengel, Y. A., (2003). Heat Transfer: A Practical Approach, 2nd ed., McGraw Hill, New York.
[26] Lenert, A. 2010.Nanofluid-based receivers for high temperature high-flux direct solar collectors, PhD thesis, Massachusetts Institute of Technology, Massachusetts.
[27] Özişik, M. N., Orlande, H. R., Colaco, M. J., & Cotta, R. M. (2017). *Finite difference methods in heat transfer*. CRC press.
[28] Versteeg, H. K., & Malalasekera, W. (2007). *An introduction to computational fluid dynamics: the finite volume method*. Pearson education.
[29] Patankar, S. V. (1980). *Numerical heat transfer and fluid flow.* Hemisphere Publishing Corporation.
[30] Teitel, M., Schwabe, D. & Gelfgat, A. Yu. (2008). Experimental and computational study of flow instabilities in a model of Czochralski growth. J. Cryst. Growth 310, 1343–1348
[31] Wolff, F., Beckermann, C., & Viskanta, R. (1988). Natural convection of liquid metals in vertical cavities. *Experimental Thermal and Fluid Science*, *1*(1), 83-91.

[32] Deshmukh, P., Mitra, S.K., Gaitonde, U.N., 2011. Investigation of natural circulation in cavities with uniform heat generation for different Prandtl number fluids, International Journal of Heat and Mass Transfer, 54, 1465–1474.